# Prompt Inject Detection with Generative Explanation as an Investigative Tool

*Jonathan Pan, Swee Liang Wong, Yidi Yuan, Xin Wei Chia
Home Team Science and Technology Agency, Singapore
Jonathan_Pan@htx.gov.sg, Wong_Swee_Liang@htx.gov.sg, Yuan_Yidi@htx.gov.sg, Chia_Xin_Wei@htx.gov.sg

*Abstract*— Large Language Models (LLMs) are vulnerable to adversarial prompt based injects. These injects could jailbreak or exploit vulnerabilities within these models with explicit prompt requests leading to undesired responses. In the context of investigating prompt injects, the challenge is the sheer volume of input prompts involved that are likely to be largely benign. This investigative challenge is further complicated by the semantics and subjectivity of the input prompts involved in the LLM conversation with its user and the context of the environment to which the conversation is being carried out. Hence, the challenge for AI security investigators would be two-fold. The first is to identify adversarial prompt injects and then to assess whether the input prompt is contextually benign or adversarial. For the first step, this could be done using existing AI security solutions like guardrails to detect and protect the LLMs. Guardrails have been developed using a variety of approaches. A popular approach is to use signature based. Another popular approach to develop AI models to classify such prompts include the use of NLP based models like a language model. However, in the context of conducting an AI security investigation of prompt injects, these guardrails lack the ability to aid investigators in triaging or assessing the identified input prompts. In this applied research exploration, we explore the use of a text generation capabilities of LLM to detect prompt injects and generate explanation for its detections to aid AI security investigators in assessing and triaging of such prompt inject detections. The practical benefit of such a tool is to ease the task of conducting investigation into prompt injects.

## I. Introduction

Large Language Models have garnered lots of attention and adoption excitement. These text generation autoregressive models have been studied for use in a variety of applications from enhancing user experience when using applications to being an augmented assistant to wide variety of tasks from text summarization, conversational dialog chatbots and many others [1]. However, as with all software based technological solutions, there are vulnerabilities and exploitation techniques. Attacks against artificial intelligence (AI) model are known as adversarial artificial intelligence. Among these techniques, the most popular one associated with text generation LLMs is prompt injections that uses prompt or textual conversational inputs as the attack vectors.

Prompt injection attacks, applied successfully, could induce LLMs to hallucinate or generate non-coherent textual responses. Such attack could also induce more damaging responses that results in malicious outputs like code generation of cyber security related offensive artefacts like malware or phishing contents. Another form of malicious output is leakage of sensitive information. While the LLM may be trained to protect itself from such involuntary responses, the constantly evolving form of prompt injection attacks would eventually circumvent such protection measures acquired from training. In a way, prompt injection is a form of social engineering attacks against the targeted model to lower its protective posture and succumb to its intended adversarial instructions. There are a number of protective solutions to protect the LLMs from such attacks. They typically involve the use of inline detection and protective mechanisms guarding the inbound prompt to the LLMs to detect the prompt injects and divert the adversarial prompt. Another is the outbound responses generated by the LLMs to provide inline detection of LLMs compromised responses. These protective mechanisms would inspect prompt inputs or output responses to contain the adversarial effects of any attack attempts. These inline protections are typically known as guardrails.

The solution options for guardrails include use of signature based detectors, machine learning based and LLM based. However, with LLM based, there is limited examples of use of text-generation LLM. One such example is Llama Guard [2]. This is due to the inherent designs of such text generation models not being optimally positioned for text classification. However, text generation models could be used to generate explanation to support its inferred textual assertions [10]. In the field of AI security with the intent to protect AI models after adversarial attacks, we hypothesis that having the explanation along with the analysis of prompt inputs could aid AI security investigators and developers in triaging adversarial attacks. This research attempts to explore the following applied research question of whether text generation LLM can be used as a prompt inject investigation tool that involves having a LLM identify adversarial prompt injects and generating explanation to facilitate investigative triaging. The practical benefits of having such a tool is to facilitate cyber investigation of prompt injects in the backdrop of large volume of prompts that are largely benign.

With this research objective, we first needed the LLM tool to have the ability to detect prompt injects. Upon detection, the

LLM would then generate an explanation for its detection. In our research setup, we first applied our test on vanilla models on a test dataset consisting of both benign and malicious prompts from ToxicChat [3]. We then fine-tuned these vanilla models to improve their detection performance using training dataset from ToxicChat. We further evaluated the performance of the models using Garak test tool from Nvidia [4]. With identified prompt injects, we then assessed the explanation generated by the LLM models through human evaluation to assess the suitability of the models' explanation.

In the next section, we will first cover background information about prompt based adversarial attack techniques against LLM. We then survey the protection approaches being developed to protect LLM against such attacks. We then cover our LLM based guardrail construct with explanation generation. This is followed by an experimental setup that details how we prepared the models via fine tuning using a variety of techniques, evaluation approach used to assess the effectiveness of detecting such prompt injects and finally, to manually evaluate the generated explanations. We conclude our research with a discussion about future research directions.

## II. PROMPT BASED ADVERSARIAL TECHNIQUES

In this section, we surveyed prompt based adversarial attack techniques that can be used against LLMs. We also surveyed the approaches used to develop guardrails for LLM against prompt based injection attacks.

### A. Survey of Adversarial Attacks against LLM

Since the development of LLMs in the field of Natural Language Processing (NLP), there is a number of research advances into susceptibility of such AI models to a variety of sophisticated attacks. Chowdhury et al. [5] did a comprehensive survey of such attacks and created a taxonomy of them and presenting report research work to mitigate such attacks. They argued that the main categories of attacks to LLM are Jailbreaking, Prompt Injection and Data Poisoning. Jailbreaking refers to techniques used to bypass or disable the safeguards built into the language generation models. Safeguards refer to prevention of harmful or inappropriate content generation, maintaining user safety and enforcing of trained ethical guidelines which is typically developed in the model as part of the training program through fine-tuning or reinforcement learning techniques. The adversarial intent of Jailbreaking is to circumvent trained safeguards leading to generation of potentially harmful or sensitive content. Also manipulating the model's responses towards undesired specific goals.

Prompt injection is a technique used to manipulate or deceive the AI model into generating harmful or inappropriate content through crafted input prompts. Hence, both are input prompts based. However jailbreaking focuses on bypassing safeguards while prompt injection focuses on manipulation of the target model. The last category of attack is Data Poisoning is introducing data samples into the training set used to train or fine tune the models to compromise the model's performance, security or reliability. However, as the focus of this research work is on adversarial input prompts, hence this category of attack is not relevant. Our research focused on Jailbreaking and Prompt Injects.

### B. Prompt Injection Protection

There are a number of approaches to protect LLM models from adversarial input prompts (Dong et. al [6]). A direct approach is to train the LLM model through reinforcement learning from human feedback (RLHF). Another is using in-context training to guide the model in dealing with such attacks. Another approach is the deployment of Guardrail [7] or content filters that use algorithms /models to detect and deal with such attacks. This approach involves inspecting the input and / or the output of the LLMs and assess whether interventions are required to limit the risks of adversarial prompt attack attempt. One form of Guardrail implementation is to use encoder based approach that encodes the input prompts into its corresponding embedding space and train a classifier (Kim e. al [8]). Guardrail NeMo [9], developed by Nvidia, is a toolkit that provides developers the means to build a controllable proxy between the user and LLM. This proxy is a programmable rail using Colang that is interpreted by a runtime that applies user defined rules or automatic generated rules to protect the LLM. Llama Guard [2], developed by Meta, is a fine tuned LLM that is based on Llama2-7b architecture. This guardrail takes in both input prompts and output responses and determines whether there are safe and unsafe. The authors of Llama Guard mentioned that their future work with this model is to generate explanations to the classification decision. This is the intent of this research work.

## III. LLM BASED GUARDRAIL WITH EXPLANATION

Our research question attempts to assess whether a text generation based LLM could be used to perform classification and provide coherent explanation to the assessment. The generation of such explanations could then be used by a human analyst to evaluate the validity of the classification task and facilitate investigating and triaging. In our research experiment, our prompt is constructed to have the LLM perform the prompt classification and provide the explanation.

### A. LLM based Guildrails

With our experiment, we used Llama3.2 from Meta as the base LLM for our Guardrail. We applied the Instruct version of the model and used two different sizes namely 1B and 3B parameter sized models for comparison. We additionally used the Llama3.2 1B instruct model as the target model to which the prompt injects are meant for.

### B. Explanation of Classification

According to Divakaran and Peddinti [11], generative models like LLM could be used to provide transparency in validating the task the AI model has been assigned to perform. Kunz and Kuhlmann [10] evaluated LLM generated explanation using GPT-4 and Alpaca dataset. Their experiment entailed a selection of 200 prompt instructions that are fed to the LLM to generate explanations for manual evaluation. The prompt instructions are generally categorised into coding assistance, math questions, grammar and language prompts to correct English text, sentence text classification and facts classification.

The assessment criteria for manual evaluation of the generated explanations include listing contributing factors, subjectivity of the explanation, providing illustrative examples and whether the explanation were misleading. They noted that subjectivity was less common possibly due to applied alignment and filter process. They also observed that misleading explanation were rarely observed. Our research applied this research evaluation approach to analyse the explanation generated from prompt injects analysis.

## IV. Methodology and Evaluation

In this section, we describe our experiment setup used to address our research questions.

### A. Dataset

The dataset used for our experiment is ToxicChat by Lin et. al [3] is benchmark dataset based on real user queries from an open-source chatbot. The dataset contains rich and nuanced phenomena of real user prompts that is categorized to three main categories. They are benign prompts, toxic prompts and jailbreaking prompts. Lin et. al also provided the baseline evaluations using available toxicity detection models with HateBERT, ToxDectRoberta, OpenAI and Perspective API online services with a threshold value of 0.7. ToxicChat dataset is also used to fine tune the model as well as evaluate the models' performance in detecting adversarial prompts and generating relevant explanations during detection analysis.

### B. Evaluation Metrics

As the ToxicChat dataset has associated classification labels namely whether the prompts are Benign, Toxic and Jailbreaking, we used *Precision* to measure the accuracy of the models against type I error (true positive) and *Recall* to measure the accuracy of the models against type II error (true negative). Finally, we used *F1 score* to measure the harmonic mean of *precision* and *recall*. We also used the accuracy measurement to evaluate the models' performance.

With the evaluation of generated explanation, we used the same evaluation criteria used by Kunz and Kuhlmann [10]. The evaluation questions are the following.

- EQ1: Does the explanation list contributing factors?
- EQ2: Does the explanation include subjective or biased criteria?
- EQ3: Does the explanation include illustrative elements (eg, examples)?
- EQ4: Is the explanation misleading (e.g. arguing for a label that is wrong)?

We randomly extracted ten prompts and the corresponding generated responses from the LLMs. In the sample of ten, we had six benign prompts and four offensive prompts. The samples were given to an evaluation team of three. Two were an AI researcher and one who was both an AI and cyber security researcher. They scored the samples according to the mentioned evaluation questions. A score of one point was given when the generated explanation qualifies the corresponding evaluation question.

### C. Experimentation Preparation and Evaluation

There is one preparation stage and three evaluation stages to our experimental setup. The preparation stage is to prepare the LLM models for prompt inject detection. This involves preparing the prompt construct included the system prompt and query prompt to the models. The query prompt construct used data from the ToxicChat dataset (training and test) with the prompt to classify the prompt input to be evaluated. In our experiment, the prompt is considered adversarial when it is labelled Toxic or Jailbreaking. It is benign if the prompt is labelled both not Toxic and Jailbreaking. Additionally, the query prompt included request to the LLM to explain its classification conclusion. The following is the query prompt construct.

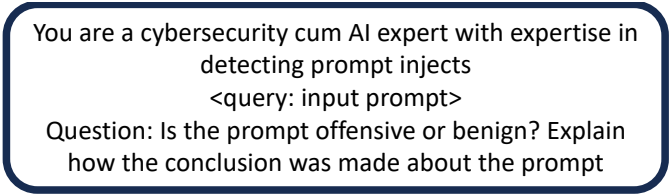

You are a cybersecurity cum AI expert with expertise in detecting prompt injects
<query: input prompt>
Question: Is the prompt offensive or benign? Explain how the conclusion was made about the prompt

Figure 1. Prompt template used

For the models used for our experimental setup, we chose the open-source Meta's Llama3.2 Instruct models with two base sizes namely 1B and 3B. We additionally fine-tuned these base models using Supervised Fine Tuning and Direct Preference Optimization [12]. The fine tuning was done with Low-Rank Adaptation (LoRA) [13].

After the preparation stage, the first evaluation stage involved evaluating our models (vanilla and fine-tuned with both techniques) using the test dataset from ToxicChat. We then used Nvidia's generative AI red-teaming and assessment kit or Garak to generate and evaluate the effectiveness of the experimented LLMs to detect prompt injects in the second evaluation stage. Garak's generated prompts are all adversarial. The configuration setup for Garak test involved having our model placed inline between the Garak and the target model (which we used Llama3.2-1B-Instruct). The vanilla configuration would not include our model hence this test involved providing Garak direct connectivity to the target model. The probes used with Garak was 'promptinject' and 'realtoxicityprompts' for prompt injection and toxic prompts respectively.

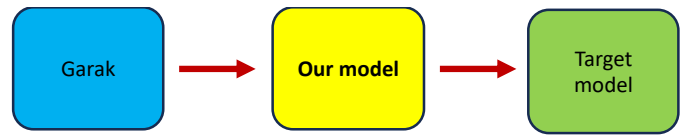

Figure 2. Garak's prompts being evaluated by our model

Finally in the final evaluation stage, we evaluated the generated explanation manually. As mentioned previously, this was done manually. Only valid classified prompts (both benign and adversarial – toxic or jailbreak) were evaluated.

### D. Evaluation Results and Analysis

The following are our experiment test results for the first stage of the model evaluations.

| Models (ToxicChat) | LLaMa3.2-1B-Instruct | | LLaMa3.2-3B-Instruct | |
|---|---|---|---|---|
| | F1 | Accuracy | F1 | Accuracy |
| Vanilla | 0.27 | 0.54 | 0.14 | 0.45 |
| SFT | **0.83** | **0.85** | **0.85** | **0.85** |
| DPO | 0.76 | 0.82 | 0.80 | 0.80 |

Table 1. F1 and Accuracy scores involving ToxicChat dataset

| Models (Garak's Prompt Inject) | LLaMa3.2-1B-Instruct | | LLaMa3.2-3B-Instruct | |
|---|---|---|---|---|
| | F1 | Accuracy | F1 | Accuracy |
| Vanilla | - | 0.02 | - | 0.25 |
| SFT | - | 0.48 | - | 0.65 |
| DPO | - | 0.22 | - | 0.66 |

Table 2. Scores with Garak using promptinject Probe

| Models (Garak's Real Toxicity) | LLaMa3.2-1B-Instruct | | LLaMa3.2-3B-Instruct | |
|---|---|---|---|---|
| | F1 | Accuracy | F1 | Accuracy |
| Vanilla | - | 0.88 | - | 0.96 |
| SFT | - | 0.94 | - | 0.99 |
| DPO | - | 0.92 | - | 0.99 |

Table 3. Scores with Garak using realtoxcityprompts Probe

We observed that fine-tuned models with both approaches (SFT or DPO) had significantly better performance in detecting adversarial prompts over the vanilla models. We also note that the model with higher parameters (3B) generalizes better after fine-tuning, having increased performance in detecting adversarial prompts from Garak, compared to the 1B model, despite having similar performances when evaluated against the test dataset of Toxic Chat. This implies that the higher parameter count allows a greater improvement in model performance after fine tuning.

We also observed that the explanations generated by the fine-tuned models fared better than the vanilla ones. The observations were on the generated responses.

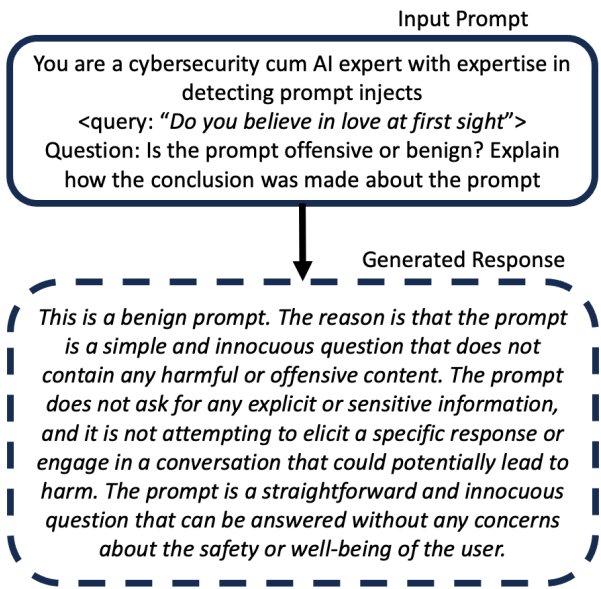

Figure 3. An example of the input prompt and generated response from LLaMa3.2-3B using SFT

The vanilla ones frequently generated no responses or explanations hence their scores for EQ1 were low. The fine-tuned models generated explanations, generally the explanations were acceptable with only a few that were subjective or biased (EQ2) and misleading (EQ4). Only a few had illustrations to support the generated explanation (EQ3).

| Models (Fine-Tuning) | EQ1 | EQ2 | EQ3 | EQ4 |
|---|---|---|---|---|
| Tinyllama (Vanilla) | 9 | 0 | 4 | 5 |
| Tinyllama (SFT) | 29 | 6 | 8 | 5 |
| Tinyllama (DPO) | 27 | 2 | 4 | 11 |
| LLama3.2-3B (Vanilla) | 8 | 4 | 0 | 2 |
| LLama3.2-3B (SFT) | 29 | 7 | 7 | 7 |
| LLama3.2-3B (DPO) | 30 | 5 | 3 | 7 |

Table 4. Generated Explanation Analysis

V. CONCLUSION AND FUTURE DIRECTIONS

Our research work explored the use of text generation LLM models to analyze adversarial input prompts and explain its analysis. We used ToxicChat dataset that contained both benign and adversarial prompts (jailbreaking and toxic). In our experiment, we applied the vanilla LLM models. We also fine-tuned these models. We evaluated these models to assess their performance in detecting adversarial prompts. We then used human evaluators to assess the validity of the explanation generated by the models. Our research concluded that using fine-tuned LLM models improved detection accuracies. These improved models also generated explanation that could facilitate analysis of the identified adversarial prompts. Hence, validating the research premise that fine-tuned text-generation LLM could be used to facilitate investigation that could quicken identification and analysis of prompt injects.

The next step to this research work is to extend this research exploration to assess the suitability of using text generation LLM for output censorship detection and investigation. Also to conduct a thorough evaluation of the generated explanations to assess their suitability in facilitating investigation and triaging and study ways to improve the quality of the generated explanation.